# Kinetic arrest, and ubiquity of interrupted $1^{st}$ order magnetic transitions.

P Chaddah *
RR Centre for Advanced Technology, Indore 452013, India

*Phase transitions are caused by varying temperature, or pressure, or magnetic field. The observation of $1^{st}$ order magneto-structural transitions has created application possibilities based on magnetoresistance, magnetocaloric effect, magnetic shape memory effect, and magneto-dielectric effect. Magnetic field induced transitions, and phase coexistence of competing magnetic phases down to the lowest temperature, gained prominence over a decade ago with theoretical models suggesting that the ground state is not homogeneous. Researchers at Indore pushed an alternative view that this phase coexistence could be due to glasslike "kinetic arrest" of a disorder-broadened first-order magnetic transition between two states with long-range magnetic order, resulting in phase coexistence down to the lowest temperatures. The CHUF (cooling and heating in unequal field) protocol created at Indore allows the observation of 'devitrification', followed by 'melting'. I show examples of measurements establishing kinetic arrest in various materials, emphasizing that glasslike arrest of $1^{st}$ order magnetic transitions may be as ubiquitous as glass formation following the arrest of $1^{st}$ order structural transitions.*

1. Introduction

First order phase transitions abound in nature; the melting of a solid and the boiling of a liquid are two of the most common phase transitions known to mankind. The conversion of liquid water into vapour, and its recondensation, correspond to first order phase transitions that sustains life; melting of ice is another first order phase transition the excess of which may threaten life on our planet. Two decades back, it was very aptly stated that physicists are guilty of giving first order phase transitions step motherly treatment! This lament may no longer be justified because of the recognition amongst material physicists that first order phase transitions are crucial to the potential for applications seen in many new materials. This has led to a resurgence of interest in first order phase transitions, with magnetic field induced transitions providing the impetus because of possible applications envisaged for magnetocaloric materials, for materials showing large magnetoresistance, for magnetic shape memory alloys, etc. Here we shall highlight the new physics that is being unfolded by experimental studies on such first order transitions.

First order phase transitions are defined by a discontinuous change in entropy at $T_C$ (resulting in a latent heat) and a discontinuous change in either volume or magnetization (depending on whether $T_C$ changes with pressure or with magnetic field). The first order phase transition can also be caused at some fixed temperature by varying a second control variable, which is either pressure or magnetic field, depending on whether the temperature driven transition is accompanied by a discontinuous change in volume or magnetic field. It is experimentally much easier to isothermally vary magnetic field than to isothermally vary pressure at temperatures well below the ambient. This is because transmission and variation of pressure requires a medium providing a thermal link which complicates the control of temperature. Magnetic field is transmitted through vacuum and its variation and control does not complicate the control of temperature. H and T induced first order transitions have a potential for applications and the ensuing extensive measurements have presented some very interesting observations. These observations, and their interpretation as originating from kinetic arrest of a first order transition, form the main theme of the paper. The idea of kinetic arrest of a disorder-broadened $1^{st}$ order transition (or a partially completed $1^{st}$ order transition being interrupted) was first mooted by us in 2001 [1] to explain anomalies observed by Tokura's group [2] in half-doped manganites (a material with potential for applications), and was justified by our extensive measurements testing its predictions in doped $CeFe_2$ (a material of purely academic interest) [1,3-6]. Magen et al [7] reported anomalies in the magnetocaloric material $Gd_5Ge_4$ which they explained using the idea of kinetic arrest [1,3,4], and they noted the similarity with half-doped manganites [2]. We subsequently made extensive studies on this material [8-11], establishing the formation of a magnetic glass and showing its devitrification. Similar anomalies were reported in Sn-doped $Mn_2Sb$ by Zhang et al [12], but were not adequately explained. We subsequently studied Co-doped $Mn_2Sb$ [13,14], and shall show some of that data, establishing kinetic arrest, here. Kinetic arrest and magnetic glass was then established by us in many different colossal magnetoresistance manganite materials [6, 15-20]. Kinetic arrest was established in NiMnIn-based magnetic shape memory alloy by Sharma et al [21], and was studied by us in NiMnSn-based alloys showing devitrification of the magnetic glass [22]. We shall present our data, showing the existence of such a kinetic arrest of a $1^{st}$ order magnetic transition, in many other materials. Prominent amongst these are Pd-doped FeRh [23] and Ta-doped $HfFe_2$ [24]. We list here reports on materials that have been recently studied only by independent groups, who have ascribed their data to kinetic arrest and formation of magnetic glass, and which shall not be covered in this paper. These newer materials identified by independent groups are Ca-doped-cobaltites [25], NiMnGa shape memory alloy [26], pure and Ba-doped samarium chromates [27], and Heusler compound $Mn_2PtGa$ [28]. The large number of materials where kinetic arrest is being reported underlines the ubiquitous nature of kinetic arrest of magnetic $1^{st}$ order transitions.

## 2. Thermodynamic metastable states: supercooling and superheating

The existence of a latent heat implies a gradual change in the fraction of transformed phase, and also that temperature cannot change from $T_C$ until the transformation is complete. The coexistence of the two phases at $T_C$ thus follows as an observation for identifying a transition as first order, as does the existence of latent heat. But first order transitions allow the existence of metastable states. At atmospheric pressure water can exist as a metastable supercooled liquid down to -40°C, and as a metastable superheated liquid to 280°C.

As depicted in the schematic solid-liquid-gas phase diagram shown in figure 1, heating a solid takes us through two first order phase transitions. We have solid-liquid coexistence at the lower transition temperature and liquid-gas coexistence at the higher transition temperature. The solid to liquid transition has some well known examples of the solid having lower density, like water, silicon, and germanium; whereas the liquid to gas transition is more generic in that the liquid has higher density in all known cases. The Van der Waal's gas provides an exactly solvable model for the first order liquid to gas transition that can be caused by varying temperature or by varying pressure. In this model the P-V isotherm is a solution of a cubic equation that corresponds to a liquid at low V and to a gas at high V. In this model the liquid changes smoothly to a gas as pressure is reduced at high T; at lower T the transformation occurs through coexistence of the two phases identifying it as a first order transition. As depicted in figure 2, below a pressure P* and above a pressure P** the cubic equation has three real roots for V. In this range of pressures one can either have a liquid, or a gas, or phase coexistence. At lease two of these are metastable states. Under thermodynamic equilibrium, at P=$P_C$ the liquid to gas phase transition occurs with V changing continuously as the coexisting fractions of liquid and gas change smoothly. This is the thermodynamically stable evolution of a Van der Waal gas. The interesting prediction is that the metastable superheated liquid can exist for reducing pressure up till P > P* and the metastable supercooled gas can exist for increasing pressure up till P < P**. However, for P* > P > $P_C$ the liquid is the stable state while for $P_C$ > P > P** the gas is the stable state.

In the Landau theory the free energy is calculated as a thermodynamic quantity that is a function of the order parameter, with the various values of the order parameter corresponding to a theoretical construct and not to observable states. This predicts limits for supercooling and superheating as a general consequence for all first order transitions, with the two limits determined by the minima in the thermodynamic free energy (vs. order parameter) reducing from two to one [29]. The Landau theory formulation explains that one can supercool or superheat only up to a limit across first order transitions, because multiple local minima in free energy exist only in a finite window T* < T < T**. However, it treats temperature as the control variable and does not treat the second control variable symmetrically. We wish to emphasize that f(S=0,T) is justifiably treated as independent of T. If, however, f is expanded in terms of a second control variable (say P), then f(S=0) cannot be assumed to be independent of P. This is obvious for the second control variable being pressure P, since change in P will change inter-atomic distances and hence the energy. This is also true for magnetic systems when the second control variable is H because varying H changes magnetization, and hence the energy. Taking this into account, Chaddah and Roy [29,30] established the important conclusion, depicted in the schematic figure 3, that the hysteresis window between supercooling and superheating encompasses a smaller T-range with rising P or H if $T_C$ rises with rising P or H, and larger T-window if it falls with rising P or H. These conclusions constituted a verifiable result. This prediction has been confirmed in many magnetic first order transitions, and a counterexample has not yet been reported. In the exactly solvable Van der Waal's model, this follows obviously from figure 2 as the hysteresis window [P*-P**] decreases monotonically with increasing T. The prediction of Chaddah and Roy depicted in figure 3 taken from [30], can be stated more generally as "When the transition is caused isothermally at T by varying the second control variable, the width of the hysteresis falls as T rises". (We note here that such isothermal measurements are experimentally difficult with pressure as the second control variable.) And "if the transition is caused by varying temperature with the second control variable held fixed, and we compare the behavior for different values of the second control variable, then the thermal hysteresis is more if the transition occurs at lower temperature".

The prediction of Chaddah and Roy is explicitly supported by studies in manganites as reported by Belvetsev et al [31], and by extensive data taken by these authors with various collaborators across many first order magnetic transitions in various materials. We show in figures 4 to 10 magnetization and resistivity measurements made as a function of T in different magnetic fields, on various materials. Hysteresis is clearly seen, and is due to supercooling and superheating across the transition. The transition temperature clearly rises (figure 4) in samples of Ta-doped $HfFe_2$ with increasing magnetic field [24] and, in accordance with the prediction of Chaddah and Roy, the hysteresis reduces. In figure 5 we show resistivity measurements on a sample of Pd-doped FeRh made in different magnetic fields [23]. The transition temperature clearly falls with increasing magnetic field and, in accordance with the prediction of Chaddah and Roy, the hysteresis increases. Similar resistivity behaviour was seen for $Mn(Co)_2Sb$ [13] and is shown in figure 6. Similar behavior but in magnetization measurements is seen for Ni-Co-Mn-Sn magnetic shape memory alloy [22] and is shown in figure 7, for $La_{0.5}Ca_{0.5}MnO_3$ [38] in figure 8, and for Ru-doped $CeFe_2$ [5] in figure 9. These are all materials in which the low-T phase has lower magnetization and $T_C$ falls with increasing H. In figure 10 we show resistivity measurements on a sample of Al-doped $Pr_{0.5}Ca_{0.5}MnO_3$ [19], which has a ferromagnetic low-T phase and where $T_C$ rises with increasing H. Again, in accordance with the prediction of Chaddah and Roy, the hysteresis reduces.

Similarly, data on isothermal transitions shows hysteresis in transition field as H is raised and lowered. The hysteresis rises when measurements are made at lower temperatures, as predicted. Data consistent with this prediction has been reported in Co-doped $Mn_2Sb$ [13], in NiMnSn-based shape-memory alloy [22], in doped samples of $CeFe_2$ [1,3], and in various manganites [15-20]. The prediction of Chaddah and Roy based on Landau theory for two control variables appears robust, as seen across many materials. To our knowledge, no counter-example has been reported.

An anomaly does, however, often appear as the transition temperature is lowered. This anomaly appears in three different forms, one of which is visible in figure 5 when the sample is cooled in H larger than 4 Tesla and the transition during cooling appears to be completed, but with an anomalous R vs. T dependence, around 50K. The transition appears to be totally (or almost totally) inhibited at values of H that would further lower the transition temperature, as is visible in each of figures 5 to 10. We shall now discuss this anomaly, and then highlight two more anomalies of a clearly qualitative nature that appear as T is lowered. We shall discuss the new concepts of kinetic arrest and magnetic glass that we introduced to explain these anomalous behaviors.

## 3. Metastable states due to "kinetic arrest"

Besides metastable supercooled and superheated states, another metastable state associated with first order transitions is that of a glass. Glasses form when kinetics of the first order transition is arrested, preserving the high-temperature structure while avoiding the first order liquid-crystal transition. Here, lack of dynamics prevails over thermodynamics. The common silicate window glass has been in the service of mankind for many centuries, and there is now an increasing awareness amongst the general public that window glass is actually a (frozen) liquid whose 'flow' occurs over astronomical time scales. Since liquids and solids have different densities, the crystallization of a liquid entails motion at a molecular level that requires non-zero time. The concept of rapidly freezing a liquid out of equilibrium has been exploited for producing splat-cooled metallic glasses whose density is that of the liquid but where kinetics has been arrested. In this context, we termed the glasslike kinetically arrested state that we established in variety of magnetic systems as "magnetic glass". The term magnetic glass describes a spatially ordered state where a first-order magnetic transition is inhibited by lack of kinetics and one obtains a glasslike arrested state. This arrested state is the high-T phase that exists at a low T where a state with competing order has a lower free energy.

The phase transitions described in schematic figures 1 and 2 occur under varying pressure P and temperature T, while H and T induced first order transitions are being extensively studied and will be the focus of this paper. The importance of the second control variable in glass formation is recognized in formation of structural glasses because the freezing point shifts with pressure, but was specifically exploited a few years ago when elemental germanium was vitrified under pressure. The glassy state was found to be retained on release of pressure [32]. The use of magnetic field as a second control parameter for arresting a first-order magnetic transition was conceived and exploited by us over the last more than ten years. The anomaly in figures 5-10 is attributed to this "kinetic arrest". We note that the anomaly is always seen as $T_C$ lowers. It is seen with decreasing H when the transition is to a ferromagnetic state with lowering temperature, and is seen with increasing H when the transition is to an anti-ferromagnetic state with lowering temperature.

## 4. Isothermal magnetic-field-induced transitions

The signature of arrest in an isothermal transition is generically quite subtle. We show in figure 11 isothermal magnetization measurements in Pd-doped FeRh, with varying H, at various T [23]. At T larger than 50K, the field for both the forward and reverse transition decreases as T rises. Below 50K, however, the field for the forward transition decreases with increasing T but the field for the reverse transition increases with increasing T. This is because the reverse transition is now caused by devitrification and not due to reaching the limit of supercooling (as was the case above 50K). The temperature dependence of the devitrification field, and that of the transition field, have opposite signs [15]. We have observed similar anomalous behavior in other materials, and show an example in figure 12 isothermal magnetization measurements in a sample of Al-doped $Pr_{0.5}Ca_{0.5}MnO_3$, which has a ferromagnetic low-T phase and where $T_C$ rises with increasing H. Here the field at which the forward transition occurs decreases with increasing T below 40K, and increases with increasing temperature above 40 K [33]. The anomalous behavior below 40K again reflects that the transformation is caused by devitrification. This non-monotonic behavior with temperature, of one of the fields at which the transition occurs in an isothermal variation, is the second qualitative anomaly.

The third qualitative anomaly was visually much more drastic, but is not observed in all materials (fortunately, we could observe it in our first report of kinetic arrest). At some low temperature we observed the visually drastic anomaly of the virgin curve (this is observed both for M vs. H and for R vs. H) jutting out of the envelope hysteresis curve. We show examples of this in R vs H measurements on $Mn(Co)_2Sb$ in figure 13 [13], in both M vs H and in R vs H measurements on $Nd_{0.5}Sr_{0.5}MnO_3$ in figure 14 [17], and in M vs H on Al-doped $Pr_{0.5}Ca_{0.5}MnO_3$ and on $La_{0.5}Ca_{0.5}MnO_3$ in figure 15 [19]. The data in figures 13 and 14 clearly shows this very visual anomaly gets enhanced as T is reduced. This anomaly arises because the ferromagnetic state obtained at large H (in $La_{0.5}Ca_{0.5}MnO_3$, in Co-doped $Mn_2Sb$, and in $Nd_{0.5}Sr_{0.5}MnO_3$) persists as a metastable kinetically arrested state even as the magnetic field is reduced to zero. Devitrification does not take place even at H=0. In Al-doped $Pr_{0.5}Ca_{0.5}MnO_3$, the zero-field-cooled state is the metastable kinetically arrested anti-ferromagnetic state that transforms to the equilibrium ferromagnetic phase with isothermal increase of H, and remains in this equilibrium phase during subsequent cycling of H. The first observation of this drastic anomaly was reported by us in Al-doped $CeFe_2$ [1], following which we introduced the concept of kinetic arrest, and used it to explain some anomalies observed by Tokura's group in half-doped manganites [2]. This anomaly was also observed in $Gd_5Ge_4$ which has a ferromagnetic low-T phase [8].

5. Cooling and heating in unequal fields (CHUF)

The schematic in figure 16 now presents our phenomenology for the formation of a magnetic glass. For a cooling field lying between $H_1$ and $H_2$ the phase transition will proceed partly, and then get arrested [15]. This results in phase coexistence down to the lowest temperature. For cooling fields outside this range, the transition is either completed or is totally arrested [15]. We predicted that by cooling and warming in different fields (that have to be chosen appropriately), one can observe a partial devitrification of the arrest state, followed by the equilibrium phase transition at higher temperature. However, the reentrant transition would be seen only for a particular sign of the inequality between the cooling field and the warming field [15,19]. And the sign for which the reentrant transition is seen would depend on whether $T_C$ rises (or falls) with increasing H; on whether the low-T equilibrium phase is ferromagnetic (or antiferromagnetic) [15,19]. This protocol that carries the acronym CHUF [19] has been exploited by us in magnetization and resistivity measurements [22-24,33] on various materials, as also by other groups recently [25,28]. We show data [19] on Al-doped $Pr_{0.5}Ca_{0.5}MnO_3$, that was used to present experimental evidence supporting our phenomenology for CHUF, in figure

17. In this material the low-T equilibrium phase is ferromagnetic, and the reentrant transition is seen only if the warming field is larger than the cooling field. In figures 18 and 19 we show CHUF measurements on $Nd_{0..5}Sr_{0.5}MnO_3$ [37] and NiMnSn-based shape-memory alloy [22] respectively. In both these materials the the low-T equilibrium phase is anti-ferromagnetic, and the reentrant transition is seen only if the warming field is lower than the cooling field.

The reentrant behavior corresponds to the states of the system evolving under the CHUF protocol, and must show up in all measurements. Accordingly, we have recently shown such reentrant behavior at a microscopic level, using neutron diffraction on NiMnSn-based magnetic shape memory alloy and on $La_{0.5}Ca_{0.5}MnO_3$ [35]. We have also used the CHUF protocol and shown reentrant behavior in Mossbauer measurements on Ta-doped $HfFe_2$ [36]. These put our phenomenology on a very firm footing, and has allowed anologous predictions when pressure (with protocol CHUP), or electric field (with protocol CHUE), is the second control variable [34].

## 6. Conclusions

In this talk I have shown examples of measurements establishing kinetic arrest in various materials, emphasizing that glasslike arrest of $1^{st}$ order magnetic transitions may be as ubiquitous as glass formation following the arrest of $1^{st}$ order structural transitions. Magnetic ordering caused by localized rather than itinerant moments could relate to the proposal of Chaddah and Banerjee [22,24,39] that a magnetic glass is formed when the magnetic latent heat is weakly coupled (cf the sample specific heat) to the thermal conduction process. Analogous predictions have been made for structural transitions where pressure would be a second control variable, and for dielectric transitions where electric field would be a second control variable [34].

## Acknowledgements

I acknowledge benefitting from long-standing collaborations with Sindhunil Roy, with Alok Banerjee, and with Rajeev Rawat. I thank all my collaborators for making such extensive measurements possible.

*Since retired; Email: chaddah.praveen@gmail.com

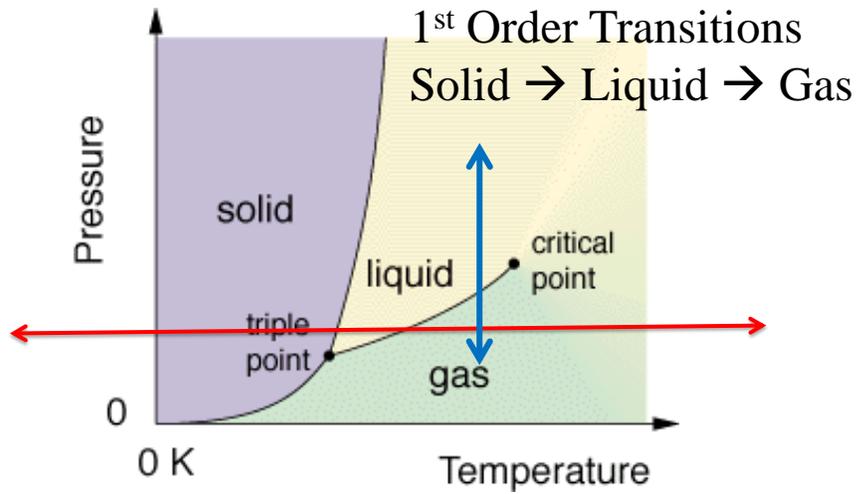

**Figure 1.** Schematic of the solid-liquid-gas 1st order transitions, that can be caused by either varying T or P. The liquid-gas phase coexistence always has a positive slope, while the solid-liquid coexistence does have a negative slope (in some range of P) for water, and for some other materials.

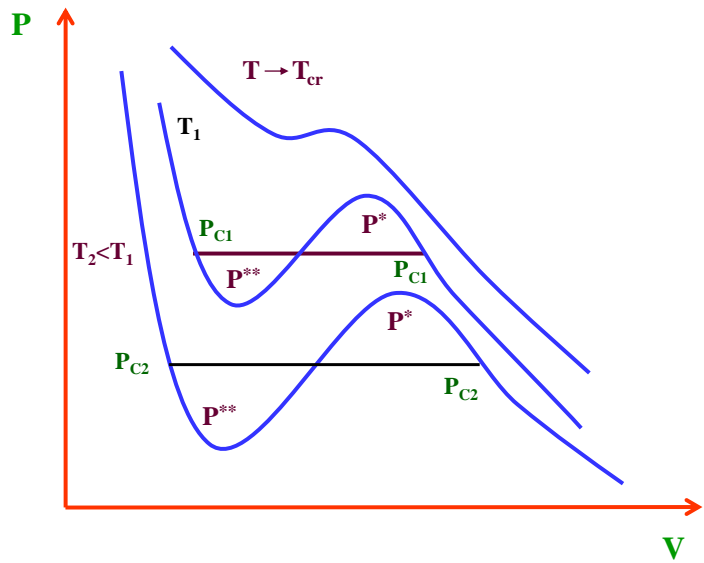

**Figure 2.** P-V isotherms for the Van der Waal's gas. Supercooling and superheating is possible.

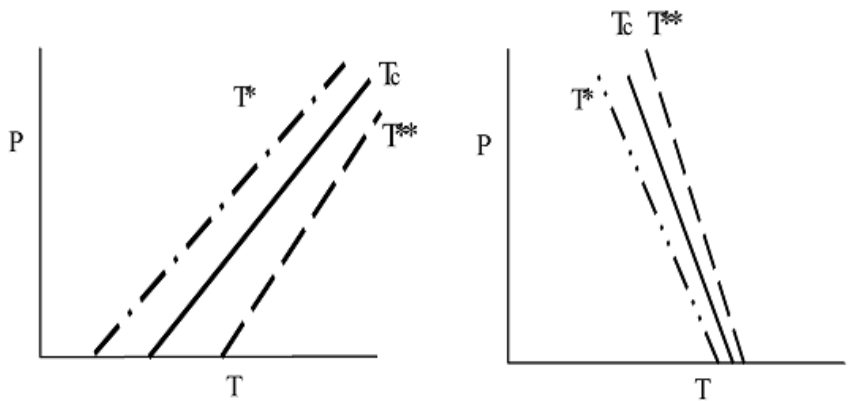

**Figure 3.** Schematic, from Chaddah [30], depicts the conjecture that hysteresis rises as the transition temperature falls.

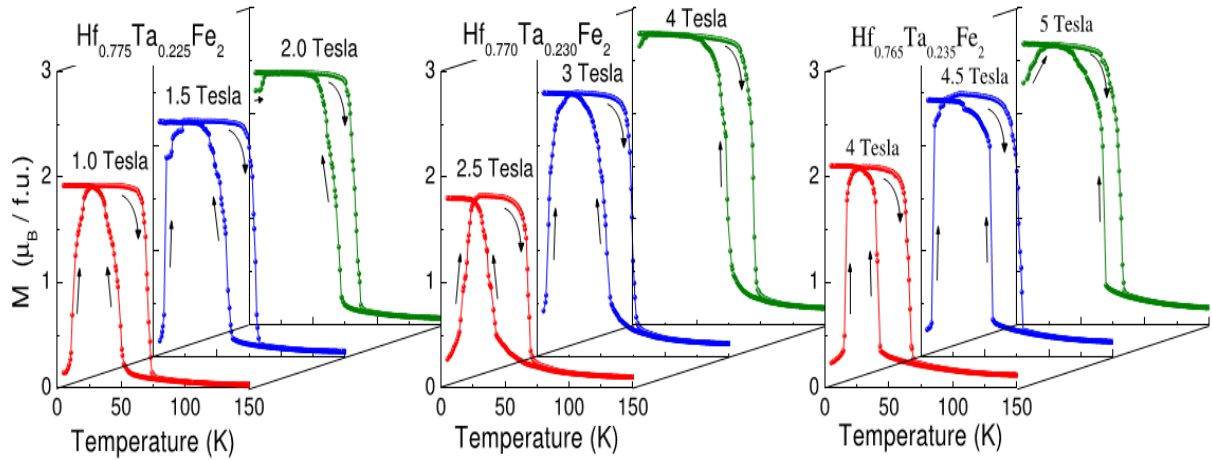

**Figure 4.** M vs T measurements, on various Ta-doped $HfFe_2$ materials, from Rawat et al [24]. Measurements are done under FCC, FCW, and ZFCW conditions in the labeled fields. For a given sample, hysteresis falls as $T_C$ rises. Also, at the lowest $T_C$, the transition under FCC remains incomplete.

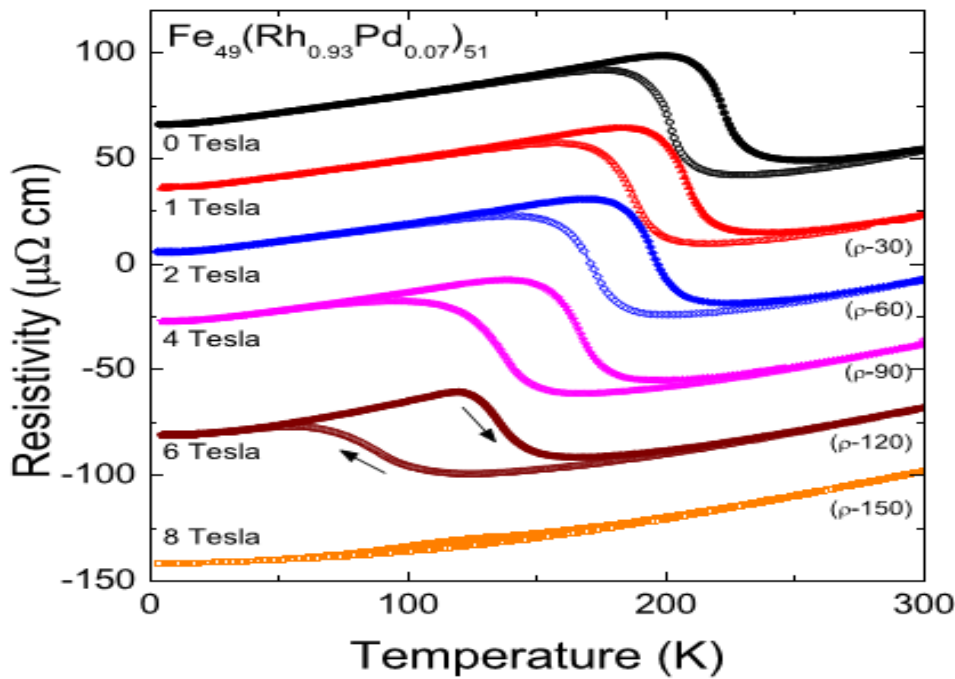

**Figure 5.** R vs. T for Pd-doped FeRh in FCC and FCW conditions, from Kushwaha et al [23]. As H rises $T_C$ drops, hysteresis increases, and the transition under FCC remains incomplete in H=6 Tesla, and is almost completely inhibited in H=8 Tesla.

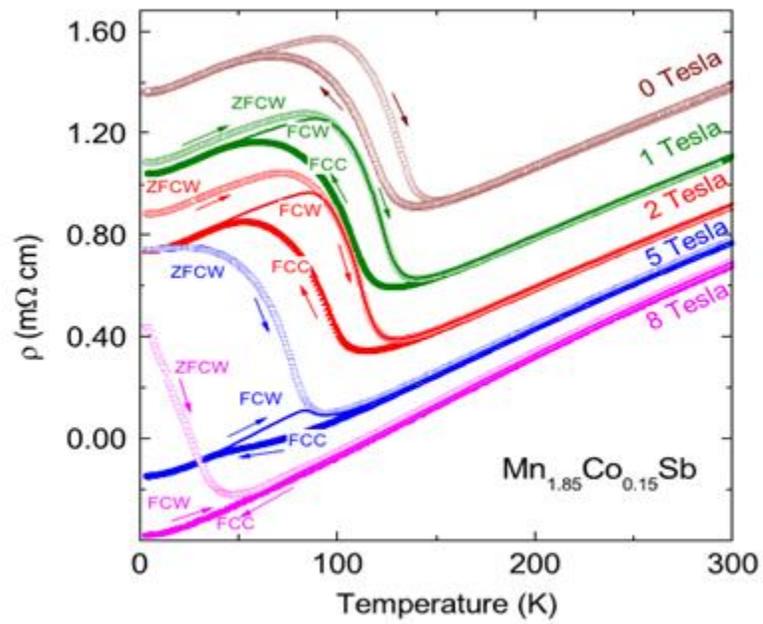

**Fig 6.** R vs. T for Co-doped Mn$_2$Sb in ZFCW, FCC and FCW conditions, from Kushwaha et al [13]. As H rises T$_C$ drops, hysteresis increases, and the transition under FCC remains incomplete in H above 2 Tesla, and is almost completely inhibited in H=8 Tesla.

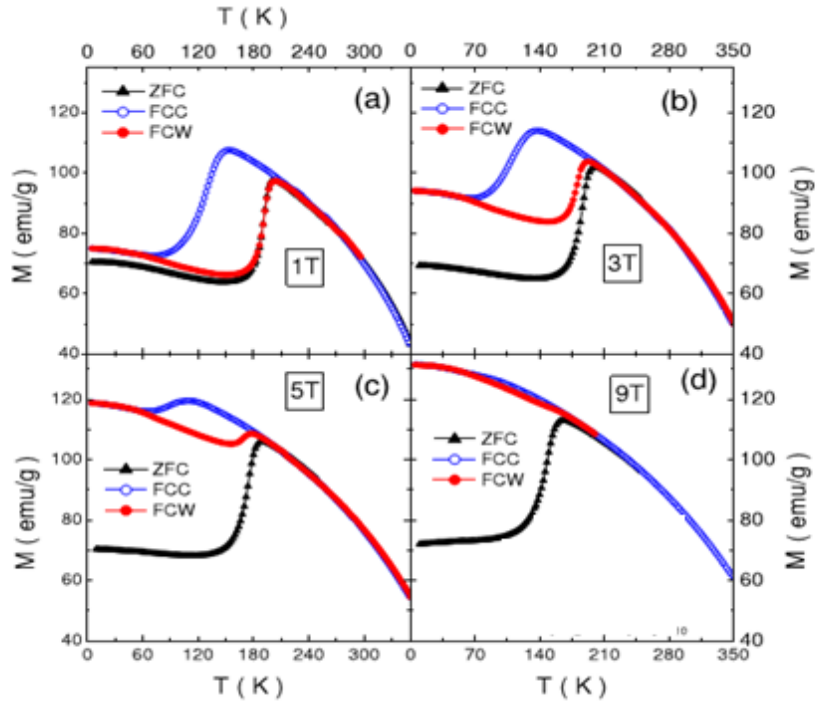

**Fig 7.** M vs. T for Co-doped NiMnSn magnetic shape memory alloy in ZFCW, FCC and FCW conditions, from Lakhani et al [22]. As H rises $T_C$ drops, hysteresis increases, and the transition under FCC remains incomplete in and above H=3 Tesla, and is almost completely inhibited in H=9 Tesla.

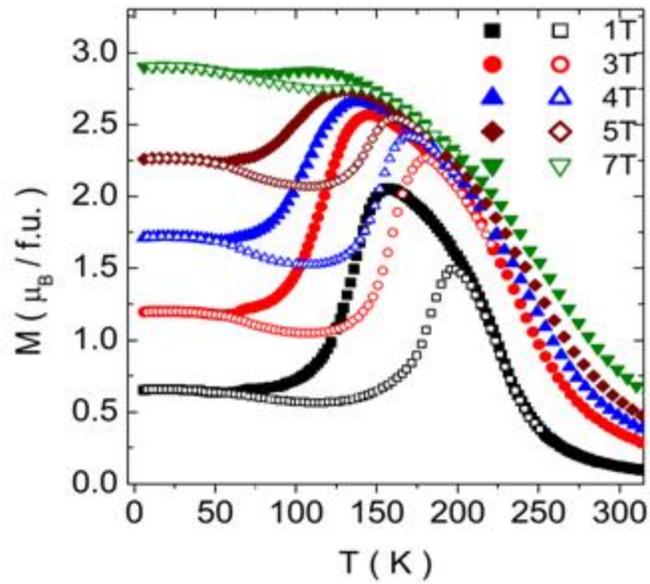

**Fig 8.** M vs. T for La$_{0.5}$Ca$_{0.5}$MnO$_3$ in FCC and FCW conditions, from Dash et al [38]. As H rises T$_C$ drops, hysteresis increases, and the transition under FCC remains incomplete in and above H=3 Tesla, and is almost completely inhibited in H=7 Tesla.

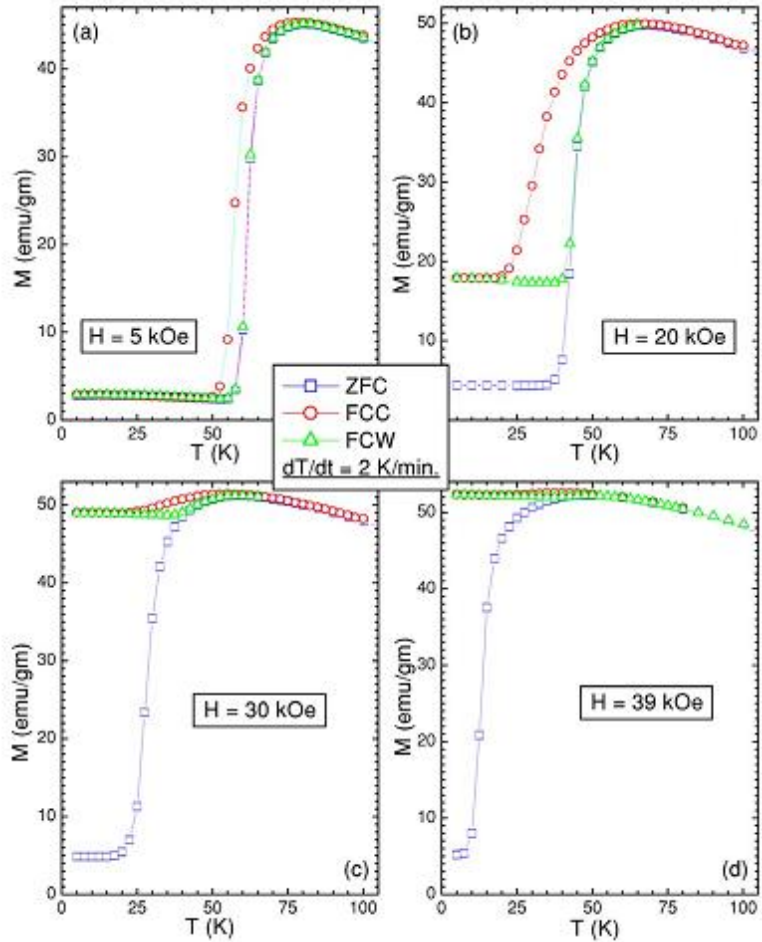

**Fig 9.** M vs. T for Ru-doped CeFe$_2$ in ZFCW, FCC and FCW conditions, from Chattopadhyay et al [5]. As H rises T$_C$ drops, hysteresis increases, and the transition under FCC remains incomplete in and above H=2 Tesla, and is almost completely inhibited in H=3.9 Tesla.

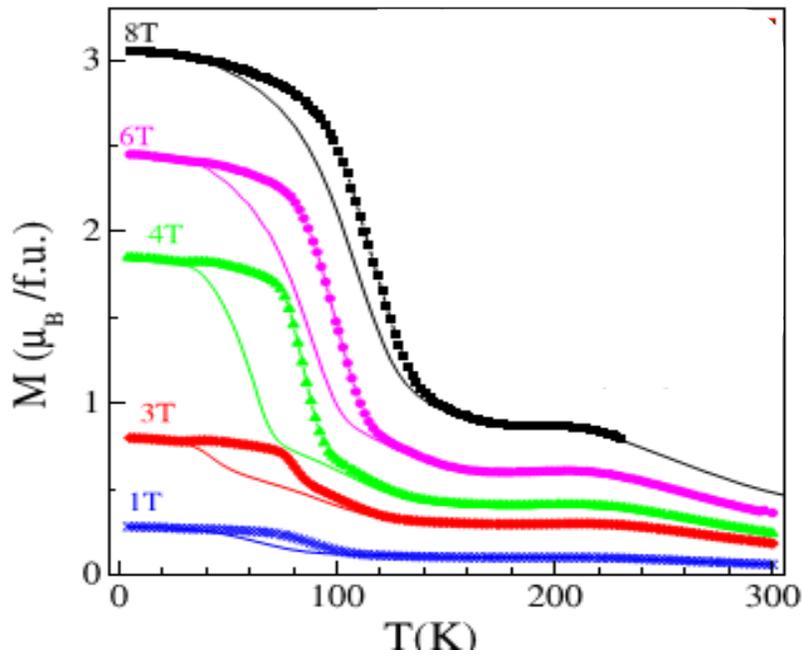

**Fig 10.** R vs. T for Al-doped $Pr_{0.5}Ca_{0.5}MnO_3$ in FCC and FCW conditions, from Banerjee et al [19]. Here the low-T equilibrium phase is ferromagnetic. As H falls $T_C$ also falls, hysteresis increases, and the transition under FCC remains incomplete in H below 4 Tesla, and is almost completely inhibited in H=1 Tesla.

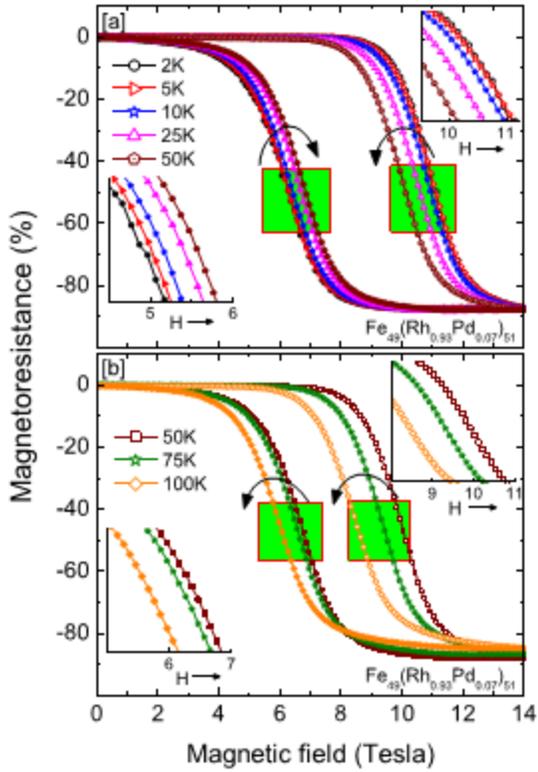

**Fig 11.** Isothermal M vs H for Pd-doped FeRh, from Kushwaha et al [23]. The value of H at which the field-decreasing transition occurs, is non-monotonic at 50K.

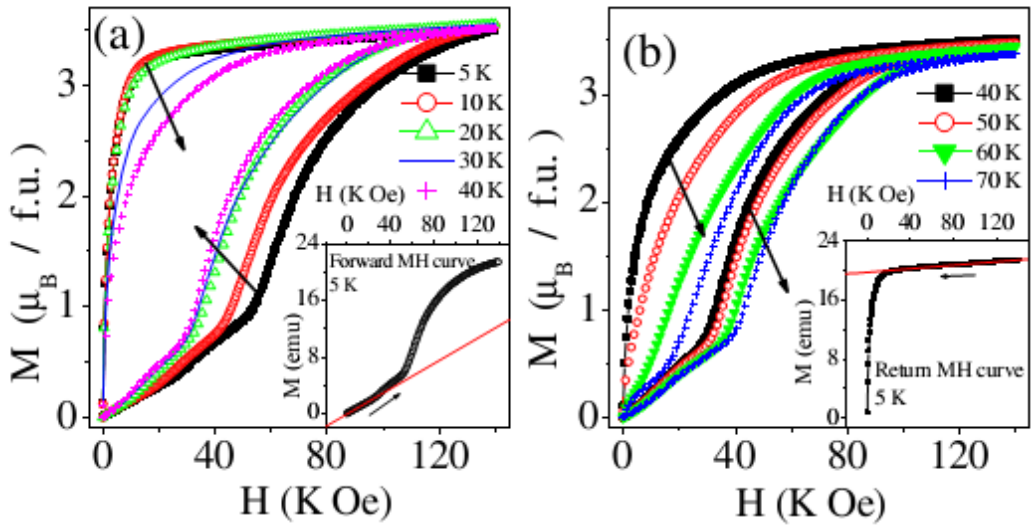

**Fig 12** Isothermal M vs H for Al-doped $Pr_{0.5}Ca_{0.5}MnO_3$, from Mukherjee et al [33]. The value of H at which the field-increasing transition occurs, is non-monotonic at 40K.

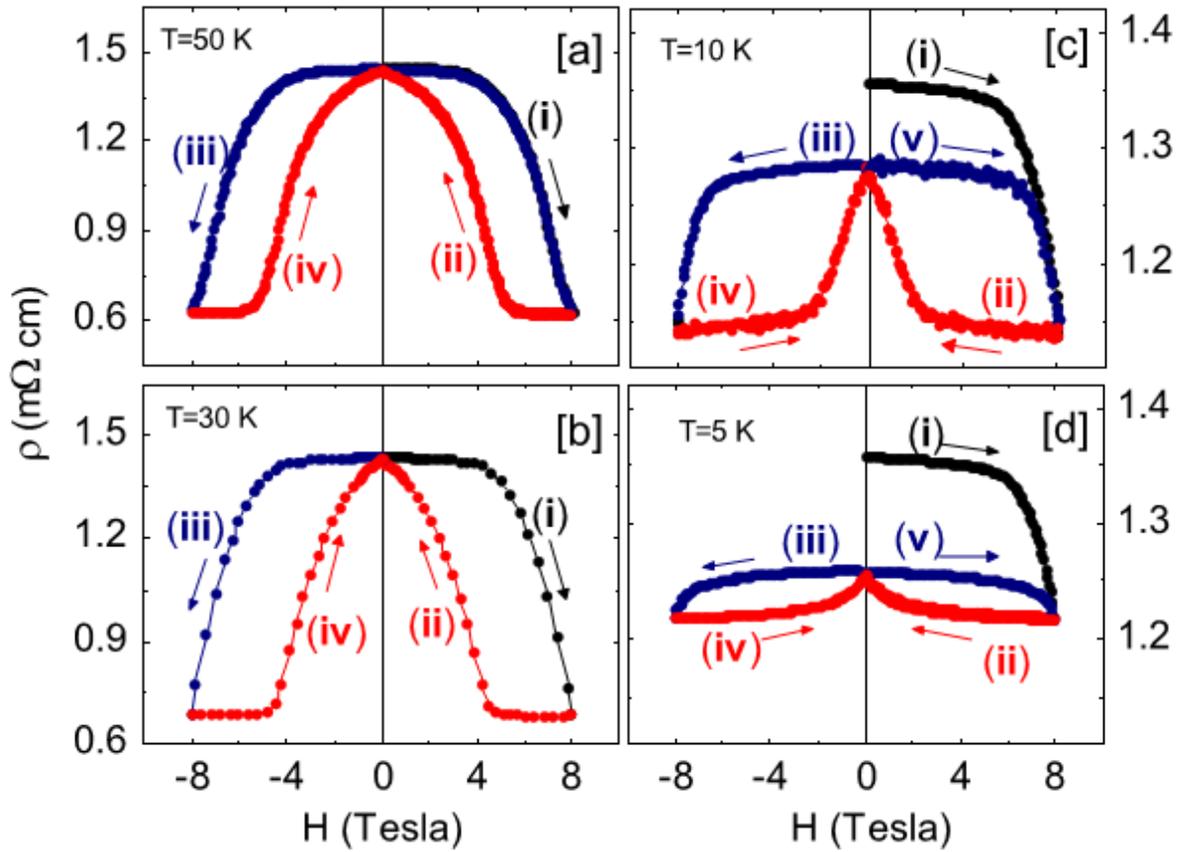

**Fig 13** Isothermal R vs H for Co-doped Mn$_2$Sb, from Kushwaha et al [13]. The virgin curve is anomalous at 10K and below, jutting out of the envelope hysteresis curve. The anomaly is more at 5K.

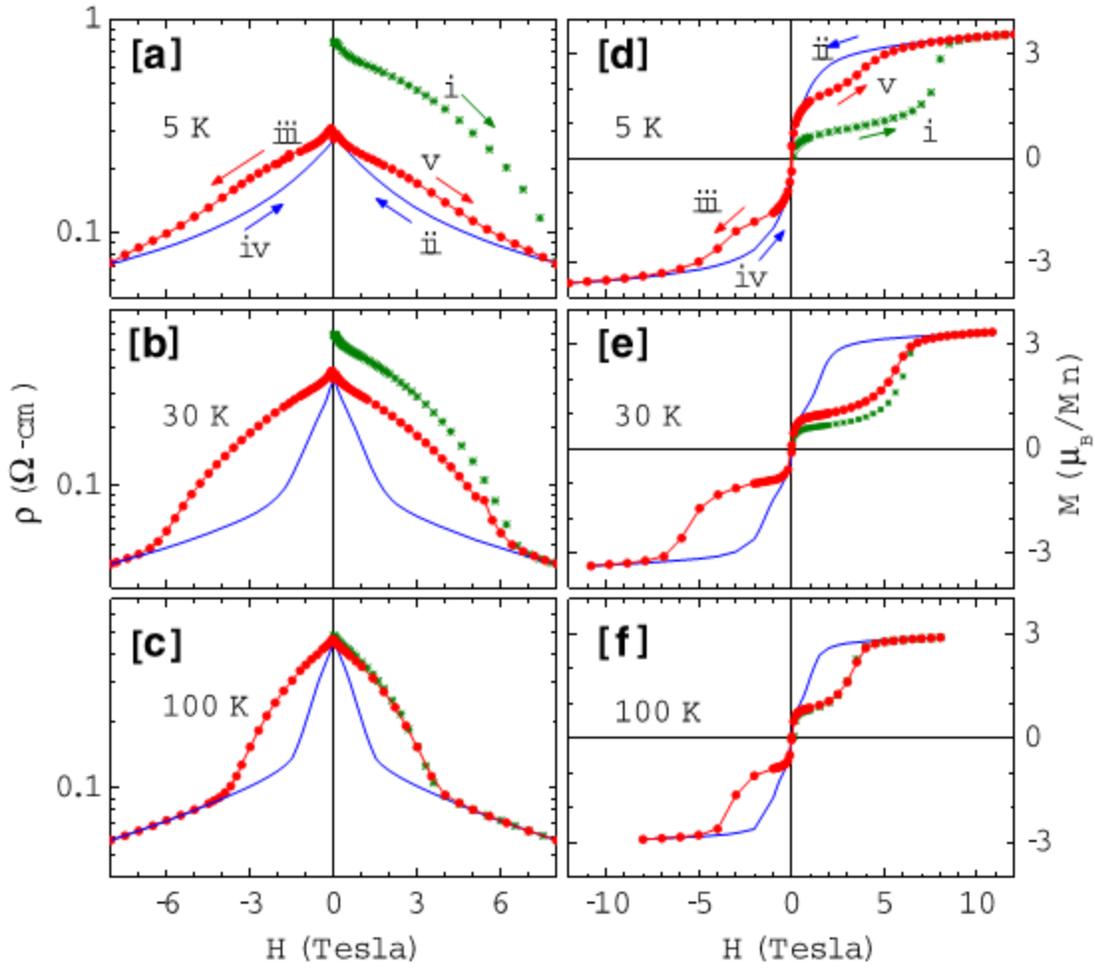

**Fig 14** Isothermal R vs H and isothermal M vs H for $Nd_{0.5}Sr_{0.5}MnO_3$, from Rawat et al [17]. The virgin curve is anomalous at 30K and below, jutting out of the envelope hysteresis curve in both measurements. The anomaly is more at 5K.

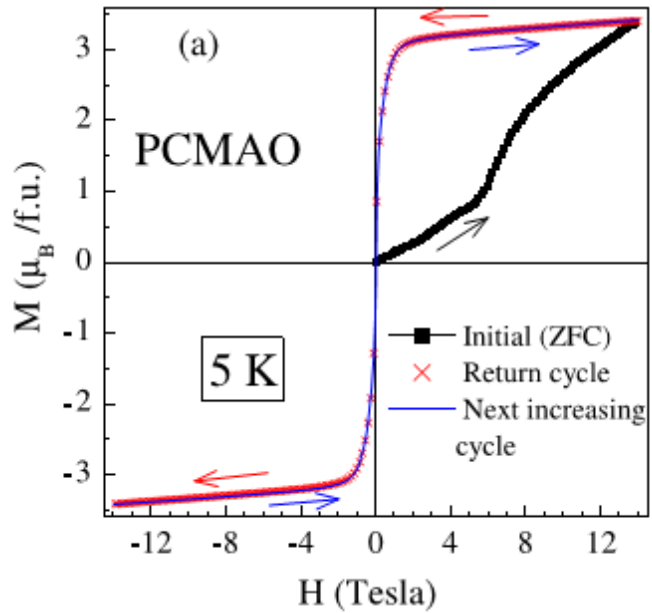

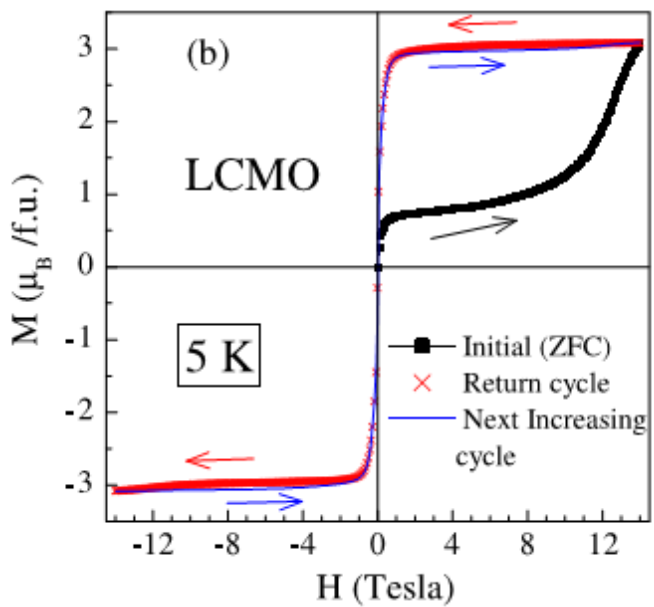

**Fig 15** Isothermal M vs H for Al-doped $Pr_{0.5}Ca_{0.5}MnO_3$, and for $La_{0..5}Ca_{0.5}MnO_3$ from Banerjee et al [19]. The virgin curve is anomalous, jutting out of the envelope hysteresis curve in both cases. The envelope curve in both cases corresponds to a soft ferromagnet, and shows no hysteresis.

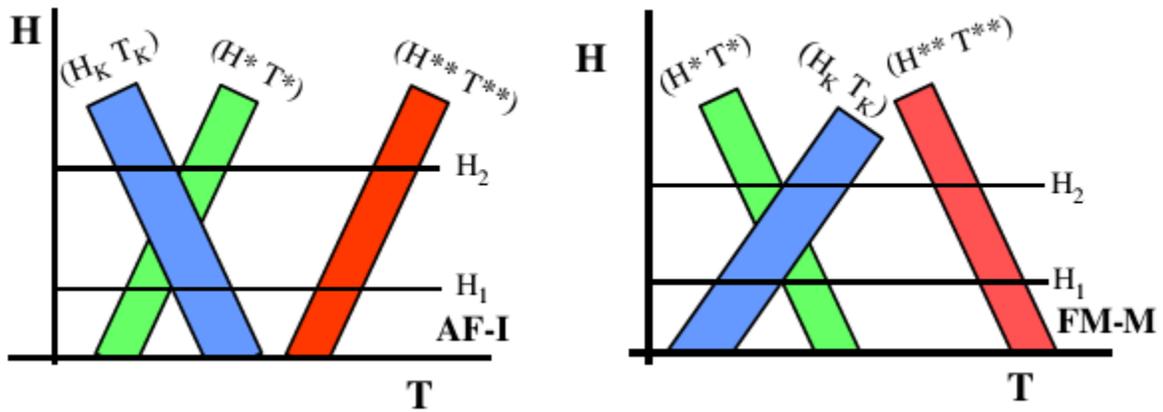

**Fig 16** The heuristic phase diagrams showing supercooling limit, superheating limit, and the kinetic arrest temperature, as a function of field, from Banerjee et al [15]. The two cases of low-T equilibrium phases are depicted. The various lines are broadened into bands due to disorder. The slopes of the bands, assumed to be constant without any loss of generality, are decided from phenomenology [15].

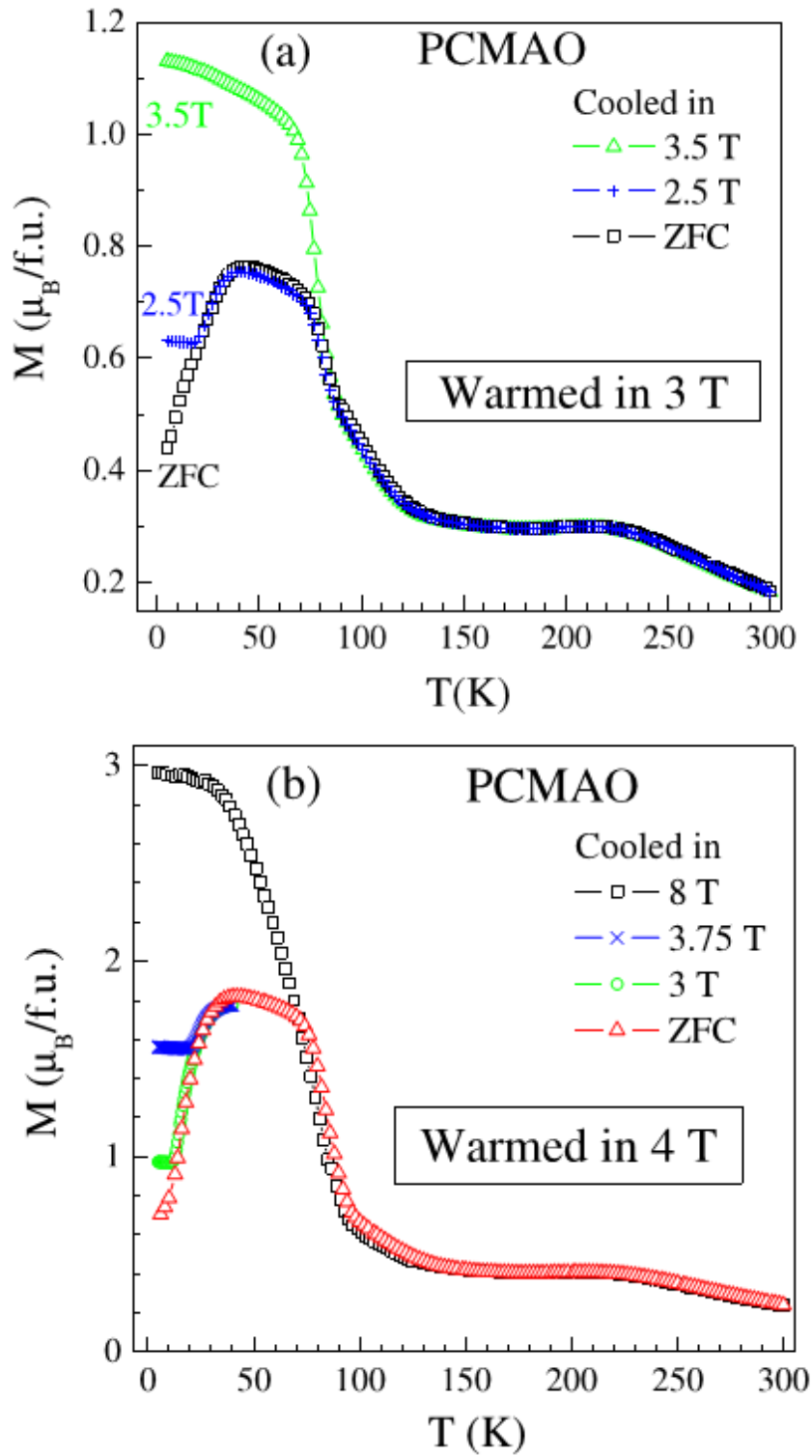

**Fig 17** Magnetization of Al-doped $Pr_{0.5}Ca_{0.5}MnO_3$ while warming, following the CHUF protocol, from Banerjee et al [19]. Two different values of warming field are shown. In both, a reentrant transition is seen when cooling field is lower than the warming field.

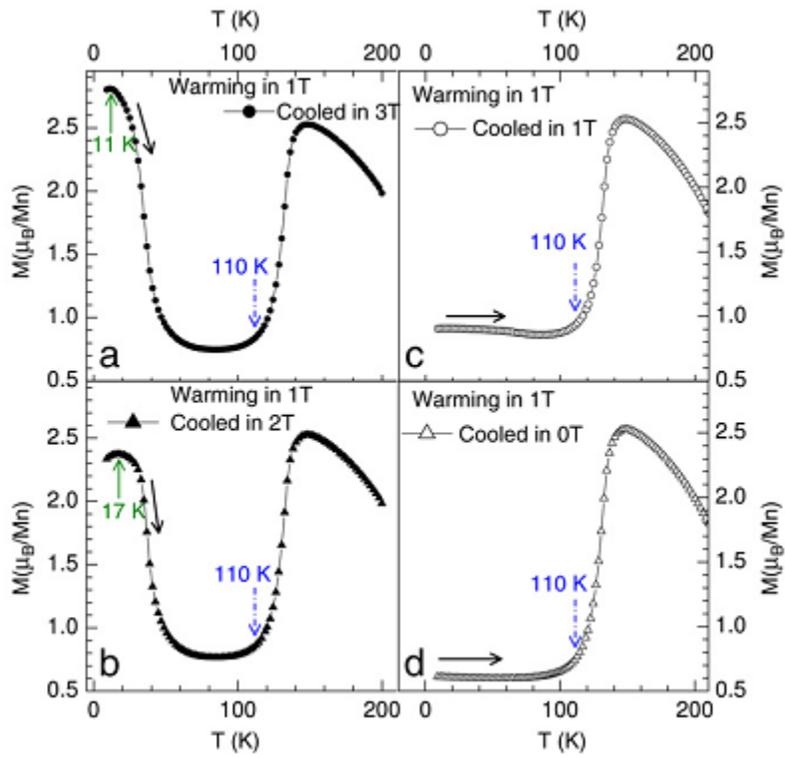

**Fig 18** Magnetization of $Nd_{0.5}Sr_{0.5}MnO_3$ measured under CHUF protocol while warming in 1 T, from Dash et al [37]. A reentrant transition is seen when the cooling field is higher than the warming field.

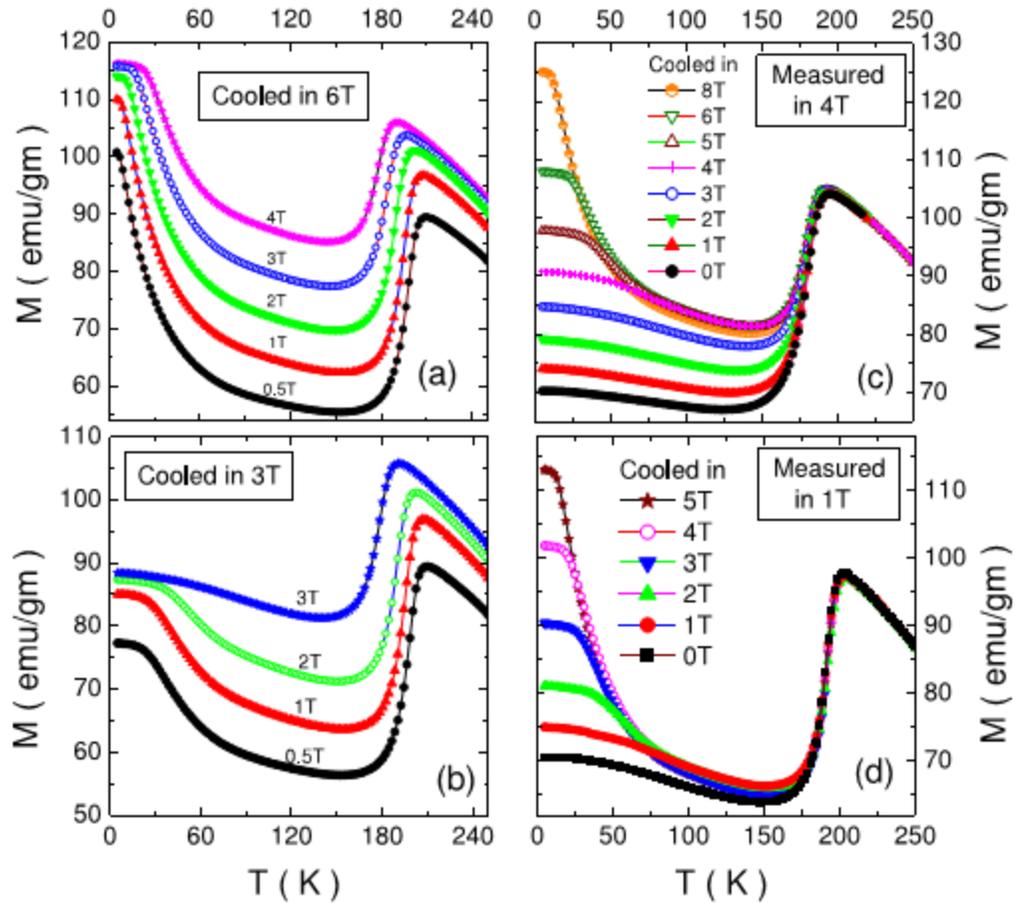

**Fig 19** Magnetization as a function of temperature using the CHUF protocol for Co-doped NiMnSn magnetic shape memory alloy, from Lakhani et al [22]. In (a) and (b) the sample is cooled in a fixed field, while in (c) and (d) the sample is warmed in a fixed field. In all cases, a reentrant transition is seen when the cooling field is higher than the warming field.